\documentstyle[aps, epsfig]{revtex}

\def\xslash#1{{\rlap{$#1$}/}}
\def\half{\frac{1}{2}}
\def\beq{\begin{equation}}
\def\eeq{\end{equation}}
\def\beqa{\begin{eqnarray}}
\def\eeqa{\end{eqnarray}}
\def\iar{\begin{array}{l}}
\def\ear{\end{array}}

\begin{document}
\draft
\title{Renormalization of the Cabibbo-Kobayashi-Maskawa Matrix at One-Loop Level}
\author{Yong Zhou$^{a}$}
\address{$^a$ Institute of Theoretical Physics, Chinese Academy of Sciences, 
         P.O. Box 2735, Beijing 100080, China}
\maketitle

\begin{abstract}
We have investigated the present renormalization prescriptions of 
Cabibbo-Kobayashi-Maskawa (CKM) matrix at one-loop level. We emphasize at one 
prescription which is formulated with reference to the case of no mixing of 
quark's generations and point out that it doesn't make the physical amplitude 
involving quark mixing ultraviolet finite. We then propose a revised prescription
to solve this problem and simultaneously keep the unitarity of the CKM matrix. 
Through explicit calculations we also prove that in order to keep the CKM matrix
gauge independent the unitarity of the CKM matrix must be preserved.
\end{abstract}

%11.10.Gh: Renormalization
%12.15.Lk: Electroweak radiative corrections 
%12.15.Hh: Determination of Kobayashi-Maskawa matrix elements
\pacs{11.10.Gh, 12.15.Lk, 12.15.Hh}

\section{Introduction}

Since the exact examination of the Cabibbo-Kobayashi-Maskawa (CKM) quark mixing 
matrix \cite{c1,c2,c3} has been developed quickly, the renormalization 
of CKM matrix becomes very important \cite{c4}. This was realized for the Cabibbo 
angle with two fermion generations by Marciano and Sirlin \cite{c5} and for the 
CKM matrix of the three-generation SM by Denner and Sack \cite{c6} more than a 
decade ago. In recent years many people have discussed this issue \cite{c7}, but 
a completely self-consistent scheme has been not obtained. In this paper we will 
study one of the present prescriptions and give some instructive conclusions.

In general, a CKM matrix renormalization prescription should satisfy the three 
criterions \cite{c8}:
\begin{enumerate}
\item In order to let the transition amplitude of any physical process involving
quark mixing ultraviolet finite, the CKM counterterm must cancel out the 
ultraviolet divergence left in the loop-corrected amplitudes. 
\item It must guarantee such transition amplitude gauge parameter independent 
\cite{c9}, which is a fundamental requirement. 
\item SM requires the bare CKM matrix $V^0$ is unitary,

\beq
\sum_k V^0_{ik} V^{0\ast}_{jk}\,=\,\delta_{ij}
\eeq
with $i,j,k$ the generation index and $\delta_{ij}$ the unit matrix element. 
If we split the bare CKM matrix element into the renormalized one and its 
counterterm 

\beq
  V^0_{ij}=V_{ij}+\delta V_{ij}
\eeq
and keep the unitarity of the renormalized CKM matrix, the unitarity of the bare 
CKM matrix at one-loop level requires 

\beq
  \sum_k (\delta V_{ik}V^{\ast}_{jk}+V_{ik}\delta V^{\ast}_{jk})\,=\,0
\eeq
\end{enumerate}

Until now there are many papers discussing this problem. The modified minimal 
subtraction ($\overline{MS}$) scheme \cite{c10,c11} is the simplest one, 
but it introduces the $\mu^2$-dependent terms which are very complicated to be 
dealt with. In the on-shell renormalization framework, the early prescription 
\cite{c6} used the $SU_L(2)$ symmetry of SM to relate the CKM counterterm with the
quark's wave-function renormalization constants (WRC) \cite{c12}. Although it is a
delicate and simple prescription, it reduces the physical amplitude involving 
quark mixing gauge dependent\footnote{This is easy to be understood since the 
$SU_L(2)$ symmetry of SM has been broken by the Higgs mechanism} 
\cite{c13,c14,c15}. A revised prescription is to replace the on-shell quark WRC
in the CKM counterterms with the ones calculated at zero momentum \cite{c13}. 
Another revised prescription \cite{c15} is to rearrange the off-diagonal quark WRC
in a manner similar to the pinch technique \cite{c16}. 

Different from the idea of Ref.\cite{c6}, another idea is to formulate the 
CKM renormalization prescription with reference to the case of no mixing of
quark's generations. This has been done in Ref.\cite{c17,c8} at one-loop level.
In the following section we will introduce this prescription and discuss the 
defect of the prescription in Ref.\cite{c17}. The revised prescription will be
given in the following section. In section 4 we will discuss the relationship 
between the unitarity and the gauge independence of the CKM matrix through 
explicit calculations. Lastly we give our conclusions.

\section{Barroso's CKM Matrix Renormalization Prescription}

The main idea of Barroso's prescription is to renormalize the transition amplitude
of W gauge boson decaying into two quarks equal to the same amplitude but in the 
case of no mixing of quark's generations. In order to elaborate this idea clearly
we firstly introduce the one-loop decaying amplitude of 
$W^{+}\rightarrow u_i \bar{d}_j$ \cite{c17}

\beqa \iar
  T_1\,=\,A_L[V_{ij}(F_L+\frac{\delta g}{g}+\half\delta Z_W +
  \half\delta \bar{Z}^{uL}_{ii}+\half\delta Z^{dL}_{jj})+
  \sum_{k\not=i}\half\delta \bar{Z}^{uL}_{ik}V_{kj}+
  \sum_{k\not=j}\half V_{ik}\delta Z^{dL}_{kj}+\delta V_{ij}]+ \\ \hspace{10mm}
  V_{ij}[A_R F_R+B_L G_L+B_R G_R]\,,
\ear \eeqa
with $g$ and $\delta g$ the $SU(2)$ coupling constant and its counterterm, 
$\delta Z_W$ the W boson WRC, $\delta\bar{Z}^{uL}$ and $\delta Z^{dL}$ the 
left-handed up-type and down-type quark's WRC \cite{c7,c18}, and

\beqa \iar
  A_L\,=\,\frac{g}{\sqrt{2}}\bar{u}_i(p_1){\xslash \varepsilon}\gamma_L 
  \nu_j(q-p_1)\,, \\
  B_L\,=\,\frac{g}{\sqrt{2}}\bar{u}_i(p_1)\frac{\varepsilon\cdot p_1}{M_W}
  \gamma_L \nu_j(q-p_1)\,.
\ear \eeqa
with $\varepsilon^{\mu}$ the W boson polarization vector, $\gamma_L$ and 
$\gamma_R$ the left-handed and right-handed chiral operators, and $M_W$ the
W boson mass. Similarly, replacing $\gamma_L$ with $\gamma_R$ in the equations we
can define $A_R$ and $B_R$ respectively. $F_{L,R}$ and $G_{L,R}$ are four form 
factors. The main idea of Ref.\cite{c17} is to choose the CKM counterterm to let
the amplitude $T_1$ similar as the amplitude of W boson decaying into two leptons.
As we know if there is no mixing of lepton's generations the amplitude of 
$W^{+}\rightarrow \nu_i \bar{e}_i$ is gauge independent and ultraviolet finite 
after introducing proper physical parameter's counterterms, except for CKM 
counterterm. So if the amplitude of $W^{+}\rightarrow u_i \bar{d}_j$ is 
renormalized equal to the same amplitude but in the case of no mixing of quark's 
generations, we will get a gauge independent and ultraviolet finite amplitude. 
This procedure will determine the CKM counterterm. The key to this problem is to
find the amplitude of $W^{+}\rightarrow u_i \bar{d}_j$ in the case of no mixing 
of quark's generations. Barroso suggested that such amplitude should be the 
following form at one-loop level \cite{c17}:

\beq
  T_1\,=\,V_{ij}[A_L(F_L+\frac{\delta g}{g}+\half\delta Z_W +
  \half\delta \bar{Z}^{uL}_{ii[l]}+\half\delta Z^{dL}_{jj[l]})+
  A_R F_R+B_L G_L+B_R G_R] 
\eeq
where the subscript "[l]" denotes the quantity is obtained by replacing CKM matrix
elements with unit matrix elements. It is easy to obtain \cite{c17}

\beq
  \delta V_{ij}\,=\,-\half\sum_k[\delta \bar{Z}^{uL}_{ik}V_{kj}+
  V_{ik}\delta Z^{dL}_{kj}]+\half V_{ij}[\delta \bar{Z}^{uL}_{ii[l]}+
  \delta Z^{dL}_{jj[l]}] 
\eeq

But in fact such CKM counterterms don't make the decaying amplitude $T_1$ 
ultraviolet finite when $i\not=j$. It is easy to calculate the ultraviolet part
of $T_1$ using $\delta V$ as CKM counterterm

$$ T_1 |_{UV-divergence}\,=\,\frac{\alpha V_{ij} \Delta}{32\pi M_W^2 s_W^2}
   (m_{d,i}^2-m_{d,j}^2+m_{u,j}^2-m_{u,i}^2) $$
with $\alpha$ the fine structure constant, $s_W$ the sine of the weak mixing angle
$\theta_W$, $m_{d,i}$ and $m_{d,j}$ the down-type quark's masses, $m_{u,i}$ and 
$m_{u,j}$ the up-type quark's masses, and 
$\Delta=2/(D-4)+\gamma_E-\ln(4\pi)+\ln(M^2_W/ \mu^2)$ (D is the space-time 
dimensionality, $\gamma_E$ is the Euler's constant and $\mu$ is an arbitrary 
mass parameter). This result shows that when $i\not=j$ the decaying amplitude of 
$W^{+}\rightarrow u_i \bar{d}_j$ is ultraviolet divergent. 

\section{One-Loop Renormalization of CKM Matrix}

We argue that such mistake comes from the one-sided knowledge about the difference
between the two cases of having and not having mixing of quark's generations. In 
the case of no mixing of quark's generations only the quarks in one generation 
can appear at a fermion line in a Feynman diagram, so only the quarks which are 
in the same generations as the external-line quarks can appear at the amplitude 
of $W^{+}\rightarrow u_i \bar{d}_j$ in the case of no mixing of quark's 
generations, except for the counterterms $\delta g$ and $\delta Z_W$. That's to 
say only two kinds of quarks: $u_i$ and $d_j$ are present in the renormalized 
amplitude of $W^{+}\rightarrow u_i \bar{d}_j$, except for $\delta g$ and 
$\delta Z_W$ because they have nothing to do with fermion. This point has been 
implicated in Ref.\cite{c8}. Thus different from Eq.(6), the amplitude $T_1$ will
be renormalized as follows

\beq
  T_1\,=\,V_{ij}[A_L(F_L+\frac{\delta g}{g}+\half\delta Z_W +
  \half \delta \bar{Z}^{uL}_{ii[l]m_{d,i}\rightarrow m_{d,j}}+
  \half \delta Z^{dL}_{jj[l]m_{u,j}\rightarrow m_{u,i}})+A_R F_R+B_L G_L+B_R G_R] 
\eeq
So the CKM counterterm is obtained compared with Eq.(4) and Eq.(8)

\beq
  \delta V_{ij}\,=\,-\half\sum_k[\delta \bar{Z}^{uL}_{ik}V_{kj}+
  V_{ik}\delta Z^{dL}_{kj}]+\half V_{ij}[
  \delta \bar{Z}^{uL}_{ii[l]m_{d,i}\rightarrow m_{d,j}}+
  \delta Z^{dL}_{jj[l]m_{u,j}\rightarrow m_{u,i}}]
\eeq
Our calculations have proven that this CKM counterterm is gauge independent and 
makes the physical amplitude $T_1$ ultraviolet convergent. 

Ref.\cite{c8} has pointed out that such a CKM counterterm maybe not satisfy the
unitary condition of Eq.(3), but an explicit Proof has not been given. Here we 
will do an analytical calculation to show in what degree $\delta V$ satisfies 
the unitary condition. At one-loop level only four diagrams need to be considered
when we calculate the CKM counterterm in Eq.(9), as shown in Fig.1.

\begin{figure}[tbh]
\begin{center}
  \epsfig{file=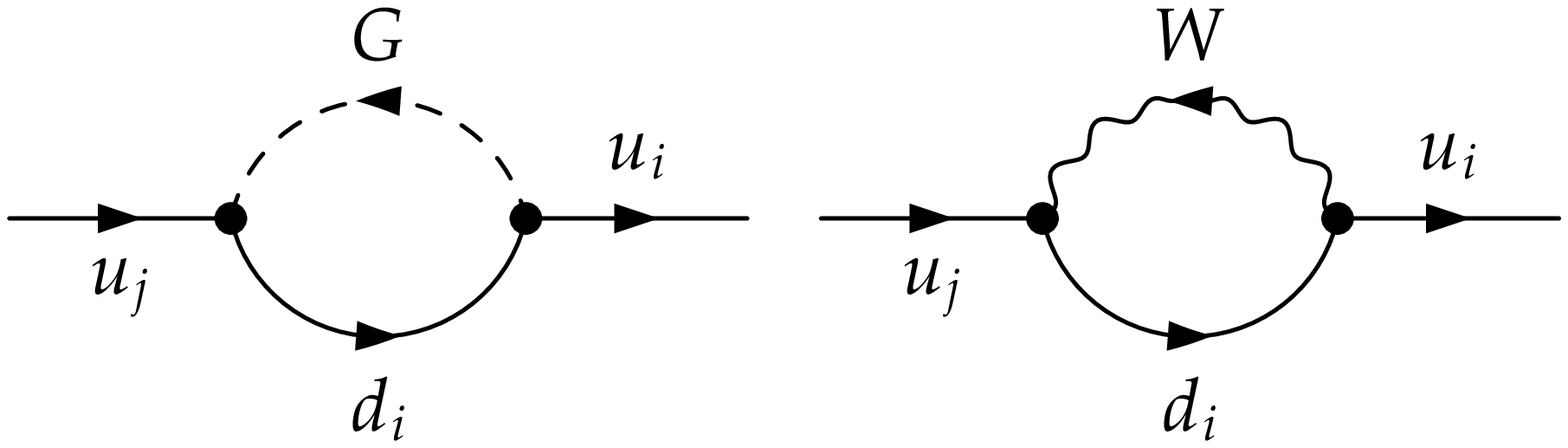, width=8cm} 
  \epsfig{file=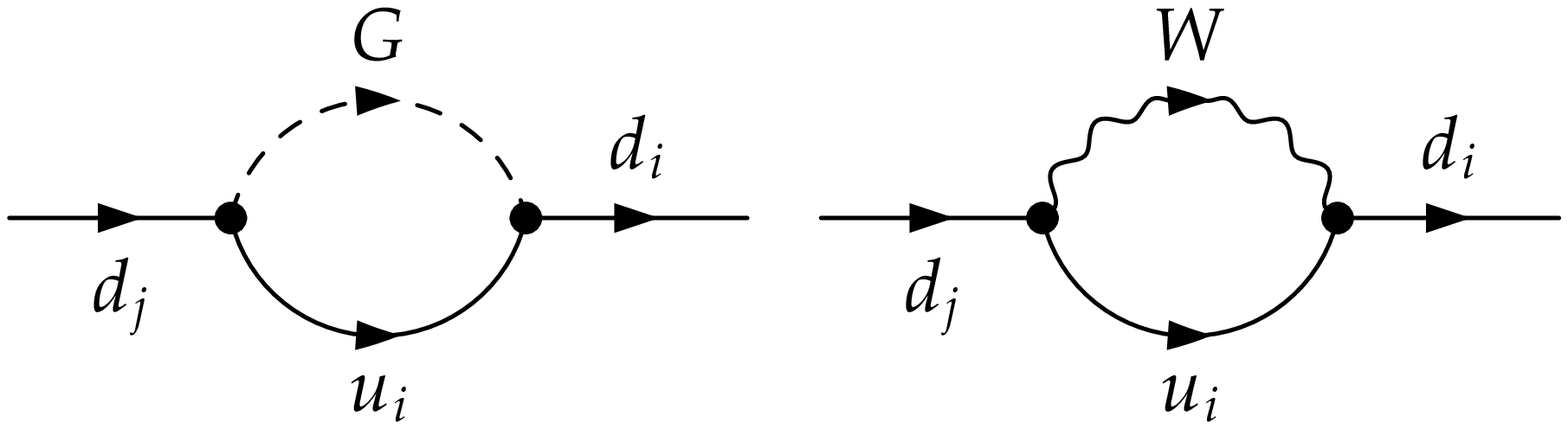, width=8cm}
  \caption{Quark's self-energy diagrams that contribute to the CKM counterterm
  in Eq.(10).}
\end{center} \end{figure}
We have used the software packages {\em FeynArts} \cite{c19} to draw the Feynman 
diagrams and generate the corresponding Feynman amplitudes, then used 
{\em FeynCalc} \cite{c20} to calculate these Feynman amplitudes. It is easy to get
the analytical result of $\delta V_{ij}$ since the quark's self-energy functions 
at one-loop level are simple. In order to check the unitary condition of Eq.(3) 
analytically, we can use the Taylor's series, $(m^2_{quark}/M^2_W)^n$, to 
expand $\delta V_{ij}$. The one- and two-order results are shown as follows:

\beqa \iar
  \hspace{-14mm}
  \delta V^{(1)}_{ij}\,=\,\frac{\alpha(6\Delta-11)}{128\pi M^2_W s^2_W}
  [-\frac{2\sum_{k,l\not=j}m_{d,j}m^2_{u,k}V_{il}V^{\ast}_{kl}V_{kj}}{m_{d,l}-
  m_{d,j}}+\frac{2\sum_{k,l}m_{d,j}m^2_{u,k}V_{il}V^{\ast}_{kl}V_{kj}}{m_{d,l}+
  m_{d,j}}-
  \frac{2\sum_{k\not=i,l}m_{u,i}m^2_{d,l}V_{il}V^{\ast}_{kl}V_{kj}}{m_{u,k}-
  m_{u,i}}+ \\  
  \frac{2\sum_{k,l}m_{u,i}m^2_{d,l}V_{il}V^{\ast}_{kl}V_{kj}}{m_{u,k}+
  m_{u,i}}+V_{ij}(\sum_k V_{ik}V^{\ast}_{ik}m^2_{d,k}+
  \sum_k V_{kj}V^{\ast}_{kj}m^2_{u,k}-2 m^2_{d,j}-2 m^2_{u,i})]\,.
\ear \eeqa

\beqa \iar
  \delta V^{(2)}_{ij}\,=\,\frac{\alpha}{64\pi M^4_W s^2_W}[V_{ij}
  (12 \ln\frac{m^2_{d,j}}{M^2_W} m^4_{d,j}+6 m^4_{d,j}+
  12 \ln\frac{m^2_{u,i}}{M^2_W} m^4_{u,i}+6 m^4_{u,i}+32 m^2_{d,j}m^2_{u,i}- 
  \\ \hspace{14mm}
  \sum_k V_{ik}V^{\ast}_{ik}(6 \ln\frac{m^2_{d,k}}{M^2_W} m^4_{d,k}+3 m^4_{d,k}+
  12 m^2_{u,i}m^2_{d,k})-\sum_k V_{kj}V^{\ast}_{kj}
  (6 \ln\frac{m^2_{u,k}}{M^2_W}m^4_{u,k}+3 m^4_{u,k}+12 m^2_{d,j}m^2_{u,k})) 
  \\ \hspace{14mm} 
  +\frac{2}{m_{d,l}-m_{d,j}}\sum_{k,l\not=j}m_{d,j}m^2_{u,k}(4 m^2_{d,j}+
  6 \ln\frac{m^2_{u,k}}{M^2_W} m^2_{u,k}+3 m^2_{u,k})V_{il}V^{\ast}_{kl}V_{kj} 
  \\ \hspace{14mm} -\frac{2}{m_{d,l}+m_{d,j}}\sum_{k,l}m_{d,j}m^2_{u,k}
  (4 m^2_{d,j}+6 \ln\frac{m^2_{u,k}}{M^2_W} m^2_{u,k}+3 m^2_{u,k})
  V_{il}V^{\ast}_{kl}V_{kj} \\ \hspace{14mm} 
  +\frac{2}{m_{u,k}-m_{u,i}}\sum_{k\not=i,l}m_{u,i}m^2_{d,l}(4 m^2_{u,i}+
  6 \ln\frac{m^2_{d,l}}{M^2_W} m^2_{d,l}+3 m^2_{d,l})V_{il}V^{\ast}_{kl}V_{kj} 
  \\ \hspace{14mm} 
  -\frac{2}{m_{u,k}+m_{u,i}}\sum_{k,l}m_{u,i}m^2_{d,l}(4 m^2_{u,i}+
  6 \ln\frac{m^2_{d,l}}{M^2_W} m^2_{d,l}+3 m^2_{d,l})V_{il}V^{\ast}_{kl}V_{kj}]\,.
\ear \eeqa
where the superscript "(1)" and "(2)" denote the one-order and two-order results
of $\delta V_{ij}$ about the series $m^2_{quark}/M^2_W$. The $R_{\xi}$-gauge 
\cite{c21} has been used. Replacing $\delta V$ with 
$\delta V^{(1)}+\delta V^{(2)}$ in Eq.(3), we find at one-loop level it satisfies
the unitary condition.  

But when we consider the three-order result of $\delta V$ about the series
$m^2_{quark}/M^2_W$, we find it doesn't satisfy the unitary condition, as shown 
below:

\beqa \iar
  \sum_k(\delta V^{(3)\ast}_{ki}V_{kj}+V^{\ast}_{ki}\delta V^{(3)}_{kj})\,=\,
  \frac{9\alpha}{128\pi M^6_W s^2_W}[\sum_k m^2_{u,k}(m^4_{d,i}-
  2 m^2_{d,j}m^2_{d,i}+m^4_{d,j}+m^2_{d,i}m^2_{u,k}+
  m^2_{d,j}m^2_{u,k})V^{\ast}_{ki}V_{kj} \\ \hspace{46mm} 
  -2\sum_{k,l,n}m^2_{u,k}m^2_{d,l}m^2_{u,n}V^{\ast}_{ki}V_{kl}V^{\ast}_{nl}V_{nj}]
  \,\not=\,0
\ear \eeqa
This result shows that $\delta V$ doesn't comply with the unitary criterion. But
from this result we can see that the deviation of $\delta V$ from the unitary 
condition is very small since the quark's masses are very small compared with 
$M_W$, except for $m_t$. Calculating to five-order result of $\delta V$ about the
series $m^2_{quark}/M^2_W$, we find the largest deviation of 
$\delta V^{\dagger}V+V^{\dagger}\delta V$ from $0$ is proportional to 
$\alpha|V_{3i\not=3}|m^2_b m^8_t/(s_W^2 M^{10}_W) \sim 10^{-7}$, which is very 
small compared with the present measurement precision of the CKM matrix elements.
Thus in actual calculations we can use Eq.(9) as the definition of the CKM 
counterterm.

Of course an severe CKM renormalization prescription should satisfy the unitary
condition severely. Diener and Kniehl have proposed a prescription which can shift
a proper CKM counterterm which has right ultraviolet divergence and gauge 
independent to satisfy the three criterions proposed in Ref.\cite{c8}, as shown
below \cite{c8}

\beq
  \delta\bar{V}_{ij}\,=\,\half(\delta V_{ij}-\sum_{k,l}V_{ik}\delta 
  V^{\dagger}_{kl}V_{lj})
\eeq
It is easy to prove that $\delta\bar{V}$ satisfies the unitary condition of 
Eq.(3). Ref.\cite{c8} has pointed out that if the ultraviolet divergence of 
$\delta V$ satisfies the unitary condition of Eq.(3) $\delta\bar{V}$ will has the
same ultraviolet divergence as $\delta V$. Because we have checked that the 
$\delta_V$ in Eq.(9) is gauge independent and contains the proper ultraviolet 
divergence which satisfies the unitary condition of Eq.(3), $\delta\bar{V}$ is 
the correct CKM counterterm at one-loop level. The explicit result is as follows

\beqa \iar
  \delta\bar{V}_{ij}\,=\,\frac{1}{4}\sum_{k}(\delta\bar{Z}_{ki}^{uL\ast}-
  \delta\bar{Z}_{ik}^{uL})V_{kj}+\frac{1}{4}\sum_{k}V_{ik}(\delta Z_{jk}^{dL\ast}
  -\delta Z_{kj}^{dL})+\frac{1}{4}V_{ij}\delta\bar{Z}_{ii[l]m_{d,i}\rightarrow 
  m_{d,j}}^{uL} \\ \hspace{12mm}
  +\frac{1}{4}V_{ij}\delta Z_{jj[l]m_{u,j}\rightarrow m_{u,i}}^{dL}
  -\frac{1}{4}\sum_{kl}V_{ik}V_{lk}^{\ast}V_{lj}(\delta\bar{Z}_{ll[l]m_{d,l}
  \rightarrow m_{d,k}}^{uL\ast}
  +\delta Z_{kk[l]m_{u,k}\rightarrow m_{u,l}}^{dL\ast})
\ear \eeqa

\section{Relationship between the Unitarity and the Gauge Independence of CKM 
Matrix}

It has been proven that using Nielsen identities \cite{c22} any physical
parameter's counterterm must be gauge independent \cite{c23}, So does the CKM 
counterterm \cite{c13}. It has been point out that in order to keep the gauge 
independence of the CKM matrix the CKM renormalization prescription must keep 
the unitarity of the CKM matrix \cite{c13,c8}. But this conclusion is only a 
hypothesis. An explicit Proof has not been given. Here we will give an explicit 
and sufficient Proof by studying the relationship between the one-loop and 
two-loop CKM counterterms.

Here one will encounter one question: if the CKM counterterms in lower-loop levels
which are gauge independent but don't satisfy the unitary condition have been 
obtained, will the CKM counterterms in higher-loop levels be gauge dependent or 
not? Our answer for this question is yes. This can been seen from the influence of
the one-loop CKM counterterm to the two-loop CKM counterterm. In order to 
elaborate this problem clearly we express the amplitude of
$W^{+}\rightarrow u_i \bar{d}_j$ in the following form

\beq
  T(V^0)\,=\,T(V+\delta V)\,=\,T(V)+T^{\prime}(V)\delta V+
  \half T^{\prime\prime}(V)(\delta V)^2+\cdot\cdot\cdot
\eeq
where the superscript $\prime$ denotes the partial derivative with respect to the
CKM matrix and the amplitude $T$ doesn't contain the CKM counterterm. To two-loop
level, this equation becomes

\beq
  T_2(V^0)\,=\,T_2(V)+T^{\prime}_1(V)\delta V_1+\delta V_2 A_L
\eeq
where the subscripts "1" and "2" denote the 1-loop and 2-loop results. 
Since $T_2(V^0)$ must be gauge independent, the gauge dependence of $\delta V_2$
is determined by the gauge dependence of $T^{\prime}_1(V)\delta V_1$ and 
$T_2(V)$. Using the facts that $F_R$ and $G_{L,R}$ are gauge independent and 
don't contain CKM matrix element and the terms in the first bracket of Eq.(4) are
gauge independent \cite{c13}, we get

\beqa \iar
  T^{\prime}_1(V)\delta V_1 |_{\xi}=[-\frac{\delta V_{ij}}{2 V_{ij}}
  (\sum_{k\not=i}\delta \bar{Z}^{uL}_{ik}V_{kj}+
  \sum_{k\not=j}V_{ik}\delta Z^{dL}_{kj})+\frac{V_{ij}}{2}\sum_{k,l}
  (\frac{2}{g}\frac{d(\delta g)}{d V_{kl}}+\frac{d(\delta Z_W)}{d V_{kl}}+
  \frac{d(\delta \bar{Z}^{uL}_{ii})}{d V_{kl}}+
  \frac{d(\delta Z^{dL}_{jj})}{d V_{kl}})\delta V_{kl}+ \\ \hspace{21mm} 
  \half(\sum_{k\not=i}\delta \bar{Z}^{uL}_{ik}\delta V_{kj}+
  \sum_{l,m,k\not=i}\frac{d(\delta \bar{Z}^{uL}_{ik})}{d V_{lm}}\delta V_{lm}V_{kj}
  +\sum_{k\not=j}\delta V_{ik}\delta Z^{dL}_{kj}+
  \sum_{l,m,k\not=j}V_{ik}\frac{d(\delta Z^{dL}_{kj})}{d V_{lm}}\delta V_{lm})]A_L
\ear \eeqa
where the subscript "1" of $\delta V_1$ has been omitted and the subscript "$\xi$" 
denotes the gauge dependent part of the quantity. Omitting the imaginary part of
the quark's self energies, we obtain

\beqa \iar
  T^{\prime}_1(V)\delta V_1 |_{\xi}=\frac{\alpha A_L}{32\pi M^2_W s^2_W m^2_{d,j}}
  \sum_{k,l}(\delta V_{il}V^{\ast}_{kl}+V_{il}\delta V^{\ast}_{kl})V_{kj}[-
  \xi^2_W M_W^4\ln\frac{m_{u,k}^2}{M_W^2}+
  2\xi_W m_{d,j}^2 M_W^2\ln\frac{m_{u,k}^2}{M_W^2}-m_{u,k}^4\ln\xi_W 
  \\ \hspace{21mm} 
  +2 m_{d,j}^2 m_{u,k}^2\ln\xi_W-2\sqrt{-\xi_W^2 M_W^4+2\xi_W m_{d,j}^2 M_W^2
  +2\xi_W m_{u,k}^2 M_W^2-m_{d,j}^4-m_{u,k}^4+2 m_{d,j}^2 m_{u,k}^2} 
  \\ \hspace{21mm}
  (\xi_W M_W^2-m_{d,j}^2+m_{u,k}^2)\arctan\frac{\sqrt{-\xi_W^2 M_W^4+
  2\xi_W m_{d,j}^2 M_W^2+2\xi_W m_{u,k}^2 M_W^2-m_{d,j}^4-m_{u,k}^4+
  2 m_{d,j}^2 m_{u,k}^2}}{-\xi_W M_W^2+m_{d,j}^2-m_{u,k}^2}]+ 
  \\ \hspace{21mm} 
  \frac{\alpha A_L}{32\pi M^2_W s^2_W m^2_{u,i}}\sum_{k,l}
  (V^{\ast}_{lk}\delta V_{lj}+\delta V^{\ast}_{lk}V_{lj})V_{ik}[-
  \xi^2_W M_W^4\ln\frac{m_{d,k}^2}{M_W^2}+
  2\xi_W m_{u,i}^2 M_W^2\ln\frac{m_{d,k}^2}{M_W^2}-m_{d,k}^4\ln\xi_W
  \\ \hspace{21mm}
  +2 m_{u,i}^2 m_{d,k}^2\ln\xi_W -2\sqrt{-\xi_W^2 M_W^4+2\xi_W m_{u,i}^2 M_W^2
  +2\xi_W m_{d,k}^2 M_W^2-m_{u,i}^4-m_{d,k}^4+2 m_{u,i}^2 m_{d,k}^2} 
  \\ \hspace{21mm}
  (\xi_W M_W^2-m_{u,i}^2+m_{d,k}^2)\arctan\frac{\sqrt{-\xi_W^2 M_W^4+
  2\xi_W m_{u,i}^2 M_W^2+2\xi_W m_{d,k}^2 M_W^2-m_{u,i}^4-m_{d,k}^4+
  2 m_{u,i}^2 m_{d,k}^2}}{-\xi_W M_W^2+m_{u,i}^2-m_{d,k}^2}]
\ear \eeqa
with $\xi_W$ the W boson gauge parameter. It is easy to be seen that if 
$\delta V_1$ satisfies the unitary condition of Eq.(3), 
$T^{\prime}_1(V)\delta V_1 |_{\xi}$ will be zero. Thus $T_2(V)$ must be gauge 
independent because in this case $\delta V_2$ is gauge independent \cite{c23}. 
On the other hand, if $\delta V_1$ doesn't satisfy the unitary condition
$T^{\prime}_1(V)\delta V_1 |_{\xi}\not=0$, $\delta V_2$ must be gauge dependent
in order to keep the amplitude $T_2(V^0)$ gauge independent.

Although this conclusion is only about the two-loop result, it is a sufficient
condition to verify the proposition: only if the unitarity of the CKM matrix is 
kept in the CKM renormalization prescription, the renormalized CKM matrix and 
its counterterm are gauge independent.

\section{Conclusion}

In summary, We have investigated the present CKM matrix renormalization 
prescriptions and found Barroso's prescription \cite{c17} has a good idea. 
But we found it doesn't make the physical amplitude involving quark mixing 
ultraviolet finite. We then modified this prescription to solve this problem and
have proven this revised prescription keep the unitarity of the CKM matrix at 
very high precision. So we can use this revised definition of the CKM counterterm
in Eq.(9) in actual calculations. A more severe CKM renormalization prescription
has been given in Ref.\cite{c8}. We also give an explicit Proof to prove that 
only if the CKM renormalization prescription keeps the unitarity of the CKM
matrix the renormalized CKM matrix and its counterterm are gauge independent.

\vspace{5mm} {\bf \Large Acknowledgments} \vspace{2mm} 

The author thanks professor Xiao-Yuan Li for his useful guidance and Dr. Hu 
qingyuan for his sincerely help (in my life). The author also thanks Prof.
B. A. Kniehl very much for his kindly criticism to help him improving this
work.

\end{document}